\documentclass{article}
\usepackage[margin=1in]{geometry}
\usepackage[utf8]{inputenc}
\usepackage[T1]{fontenc}    % use 8-bit T1 fonts
\usepackage{hyperref}       % hyperlinks
\usepackage{color}
\usepackage[super]{natbib}
\usepackage{url}            % simple URL typesetting
\usepackage{booktabs}       % professional-quality tables
\usepackage{amsfonts}       % blackboard math symbols
\usepackage{amsmath}        % for math symbols 
\usepackage{physics}
\usepackage{nicefrac}       % compact symbols for 1/2, etc.

\usepackage{authblk}        % comment out for jpc
\usepackage{microtype}      % comment out for jpc

\usepackage{graphicx}
\usepackage{subcaption}
\usepackage{lmodern}       % comment out for jpc

\def\cD{{\cal D}}
\def\cE{{\cal E}}
\def\cF{{\cal F}}

\def\vr{\mathbf{r}}

\def\vd{\mathbf{d}}

\newcommand{\mb}[1]{\mathbf{#1}}
\newcommand{\mr}[1]{\mathrm{#1}}

\title{Ground state energy functional with Hartree-Fock efficiency and chemical accuracy}
\author[a]{Yixiao Chen}
\author[a]{Linfeng Zhang}
\author[b]{Han Wang\thanks{wang$\_$han@iapcm.ac.cn}}
\author[a,c]{Weinan E\thanks{weinan@math.princeton.edu}}
\affil[a]{Program in Applied and Computational Mathematics, Princeton University, Princeton, NJ, USA}
\affil[b]{Laboratory of Computational Physics, Institute of Applied Physics and Computational Mathematics, Huayuan Road 6, Beijing 100088, People's Republic of China}
\affil[c]{Department of Mathematics, Princeton University, Princeton, NJ, USA}

\begin{document}

\maketitle

\begin{abstract}
We introduce the Deep Post–Hartree–Fock (DeePHF) method, a machine learning-based scheme for constructing  accurate and transferable models for the ground-state energy of electronic structure problems. 
DeePHF predicts the energy difference between results of highly accurate models such as the coupled cluster method and low accuracy models such as the the Hartree-Fock (HF) method,
%and density functional theory (DFT),  
using the ground-state electronic orbitals as the input.
It preserves all the symmetries of the original high accuracy model.
The added computational cost is less than that of the reference HF or DFT and scales linearly with respect to system size. 
% \WE{here we should also say something about universality.}
We examine the performance of DeePHF on organic molecular systems  using publicly available datasets and obtain the state-of-art performance, particularly on large datasets.
\end{abstract}

\section{Introduction}
Predicting the ground-state energy of a many-electron system in an environment of clamped ions is one of the most important problems in  quantum chemistry.
There is a well-known trade-off between efficiency and accuracy.
Low level models, such as density functional theory (DFT)~\cite{kohn1965self} and Hartree-Fock (HF)~\cite{hartree1935self}, 
are quite efficient but their accuracy is often less than adequate.
%, do not give good energy predictions. 
Higher-order schemes based on, e.g., the M{\o}ller-Plesset perturbation~\cite{moller1934note}, coupled cluster~\cite{vcivzek1966correlation}, configuration interaction~\cite{pople1987QCI}, or adoption of multiple references~\cite{jeziorski1981coupled}, can generate much more accurate energies, but their much increased computational expense limits their application to no more than dozens of electrons. 

In recent years,  machine learning (ML) methods have brought some new hope in this difficult area.
Significant progress has been made by using ML-based models to represent the ground-state electron energy directly as a function of the positions of the
ions and their chemical species~\cite{behler2007generalized,bartok2010gaussian,rupp2012fast,chmiela2017machine,schutt2017schnet,smith2017ani,han2017deep,zhang2018deep,zhang2018end}.
Combined with the state-of-art high performance capabilities, molecular dynamics simulations of systems of up to 100 million atoms have been performed, with an accuracy comparable to that of electronic structure models~\cite{lu2020hpc}.
However, such atom-based ML models usually do not transfer well between different chemical environment. 
At the same time, the amount of training data required, in terms of the number of atomic configurations and the system size, 
is  beyond the current capability of high level methods such as CCSD(T).

The recently developed molecular-orbital-based machine learning (MOB-ML) method \cite{welborn2018transferability,cheng2019universal,cheng2019regression} has followed a different route. The idea is to fit directly the post-HF correlation energy using information from the electronic orbitals from HF solutions. Since these models do not rely on any atomic-based features, 
they exhibit much better transferability, with only a small amount of training data.

In this work, we propose an alternative approach in the spirit of the MOB-ML scheme. 
We suggest a systematic way of devising features from the HF ground state electronic orbitals,
and build our model so that the commonly desired properties including universality, locality, symmetry
are all satisfactorily addressed.
We call this methodology DeePHF, standing for Deep Post Hartree-Fock method.
We also propose an active learning procedure which enables us to develop a 
uniformly accurate model with a minimal set of training data.
Our model exhibits the up-to-date best performance on the same set of benchmark tests used in previous work~\cite{cheng2019universal,cheng2019regression,dick2020machine}.
 
The scheme proposed here also serves as a preliminary step for our next objective: Developing systematic generalized Kohn-Sham models with uniform chemical accuracy. 
A key ingredient there is the exchange-correlation (XC) functional.  
The scheme proposed here is developed for that purpose. %~\cite{chen2020deepxc}.
 
\section{Methods}
\subsection{Problem setup}
Consider a system containing $N$ electrons indexed by  $i$ and $M$ clamped ions indexed by $I$. 
Our starting point is the Hartree-Fock (HF) orbitals $\{\ket{\psi_i}\}$, which are solutions of
\begin{equation}
    \label{eqn:HF} 
    \hat{H}_{\mr{HF}} \ket{\psi_i} = \varepsilon_i \ket{\psi_i};~\braket{\psi_i}{\psi_j}=\delta_{ij},
\end{equation}
with the HF Hamiltonian $\hat{H}_\mr{HF}$ defined as
\begin{equation}
    \hat{H}_\mr{HF} = \hat{H}_0 + \sum_i^N \hat{J}_i + \sum_i^N \hat{K}_i.
\end{equation}
%and the KS Hamiltonian $\hat{H}_\mr{KS}$  defined as
%\begin{equation}
%    \hat{H}_\mr{KS} = \hat{H}_0 + \sum_i^N \hat{J}_i + %\hat{V}_\mr{XC}.
%\end{equation}
Here $\hat{H}_0 = -\frac{1}{2}  \nabla^2 + \sum_I \frac{Z_I}{|R_I - r|}$ 
is the single electron non-interacting Hamiltonian,
%$\hat{J}_i$, $\hat{K}_i$, and $\hat{V}_\mr{XC}$ denote the Coulomb operator, the exchange operator, and the XC potential, respectively. 
$\hat{J}_i$ is the Coulomb operator, and $\hat{K}_i$ denotes the exchange operator.
For simplicity we have also used $i$ to index the orbitals.
Note that the full set of solutions $\{\ket{\psi_i}\}$ to Eq.~\ref{eqn:HF} is a complete basis set and spans the full Hilbert space. The eigenvalues $\{\varepsilon_i\}$ are real and ordered non-decreasingly, so orbitals with $i\leq N$ are  occupied orbitals; while the unoccupied orbitals, or virtual orbitals, are indexed by $i>N$.

Our goal is to model  the energy difference between CCSD(T) and HF models $E_c = E_\mr{EXACT} - E_\mr{HF}$ as a function of  the
%The original input of our model should be the solution of the HF/KS equation, namely the self-consistently obtaine
 HF single-electron orbitals $\{\ket{\psi_i}\}$:
%Therefore, essentially our target is to learn a functional 
$$E_c\equiv E_c[\{\ket{\psi_i}\}]$$
In the context of HF, $E_c$ is nothing but the correlation energy.
The existence of such a functional is an obvious fact for the quantum chemistry community.
Our objective is to construct an accurate and transferable approximation to this functional.
For this work, we will define  $E_\mr{EXACT} $ as  $ E_\mr{CCSD(T)} $.
%, but the order of expansion and the corresponding convergence rate constitute the main difference among different methods.
% $E_c$ is  the correlation energy in the context of HF.

Ideally we would like our model to have the following features.
%From the viewpoint of modelling, the model has to satisfy several nice properties:
\begin{enumerate}
\item {\it Universality}. The model should be universally applicable for a large range of systems.
It should be noted that at this point, we are not aiming at true universality. Instead we take a more programmatic approach
and aim for models that are applicable for all the systems whose local electronic configurations are well represented by the training data.
For this purpose, the input of the model should be purely electronic data, e.g., no information about the
chemical species should be used.
\item {\it Locality}. The model should be relatively local, so that it can potentially
 be constructed using data from small systems and is generalizable to larger ones.
\item {\it Symmetry}. The model should respect both physical and gauge symmetries.
% and should be invariant under corresponding transformations.
Here physical symmetry means that $E_c$ should be invariant under translation and rotation of $R_I$ as well as permutation of the ionic indices for the same chemical species.
Gauge symmetry means that $E_c$ should be invariant when the occupied orbitals $\{\ket{\psi_i}\}$ undergo a unitary transformation.
\item {\it Accuracy}. The model should at least achieve  chemical accuracy, i.e.~a prediction error lower than 1 kcal/mol.
%  \WE{is this an accurate statement?}
\item {\it Efficiency}. The cost of solving the model should be comparable to that of  HF
models.
\end{enumerate}

%From the viewpoint of learning, due to the huge cost to calculate $E_c$, we have to devise a procedure to develop a uniformly accurate model with a minimal yet adequate set of training data. 
%This is a criterion typically missed by the molecular modeling community when ML methods are adopted.
%However, since ML methods involve much more parameters than typical empirical electronic structure model, the training scheme, especially the sample of training data, is crucial to generate a truly reliable model.

\subsection{The energy model}
%To achieve our goal and satisfy the requirements listed above, 
We draw inspirations from the density (matrix) functional theory. 
It was proved  decades ago that the ground state energy is a universal functional of the electron density $n(x)$~\cite{hohenberg1964inhomogeneous}. 
Therefore it is also a unique functional of the one particle density matrix $\Gamma(x,x')$ \cite{gilbert1975hohenberg}. 
Following the same procedure as for the density functional theory \cite{kohn1965self}, for the corresponding non-interacting system we
define the density matrix as 
\begin{equation}
    \label{eqn:density-mat}
\Gamma(x,x') = \sum_i^N \bra{x}\ket{\psi_i}\bra{\psi_i}\ket{x'} = \sum_i^N \psi_i(x)\psi_i^*(x'). 
\end{equation}
It is straightforward to see that $\Gamma(x,x')$ is invariant under gauge transformation.
The correlation energy $E_c$ is therefore also a  functional of the density matrix $\Gamma(x,x')$.

% We find that it would be helpful to comment briefly on the decision to use  $\Gamma(x,x')$ as our starting point.
% Density matrix gives us more information than pure density. 
% However, it is  not clear how local this functional dependence is.
% For example, terms explicitly involving virtual orbitals like those in typical post-HF models have been dropped. 
% Our strategy is to use $\Gamma(x,x')$ as a starting point for 
% %or the presentation of our essential considerations 
% representing  $E_c$  and our experience has shown that this representation does 
% %constructed with $\Gamma(x,x')$ 
% show satisfactory expressive power and transferability. 
% Adoption of explicit information from virtual orbitals should be straightforward if needed.
% Moreover, for isolated systems like molecules and insulating systems in condensed phase, we have $\Gamma(x,x')\sim\exp({-a|x-x'|})$ when $|x-x'|$ is large~\cite{benzi2013decay},
% i.e. $\Gamma(x,x')$ is indeed localized. 
% In fact, Lin et al. used this to develop the Selected Column of Density Matrix (SCDM) method, an efficient and robust way for computing localized orbitals~\cite{damle2015compressed} and this has been generalized to metallic cases~\cite{damle2018disentanglement}.

Next we introduce a set of atomic bases centered on each atom $I$, denoted as $\left\{\ket{\alpha^I_{nlm}}\right\}$: 
\begin{equation}
    \braket{\vr}{\alpha^I_{nlm}} = R^I_n(r) Y^I_{lm} (\theta, \phi),
\end{equation}
where $R_n(r)$ is a radial function indexed by $n$, $Y_{lm}$ denotes the spherical harmonics  of degree $l$ and order $m$, 
and we use $(r,\theta, \phi)$ to denote the spherical coordinates relative to atom $I$.
For each atom we project every orbital onto the basis centered at that atom to obtain a set of overlap coefficients
\begin{equation}\label{eqn:c}
    c^I_{inlm} = \braket{\alpha^I_{nlm}}{\psi_i}.
\end{equation}
Note that in practice we will have $\ket{\psi_i} = \sum_a \lambda_{ia}\ket{\chi_a}$, where $\{\ket{\chi_a}\}$ 
is the basis set  used to perform the HF  calculation, 
 $\{\lambda_{ia} \}$ are the corresponding expansion coefficients. Therefore $c^I_{inlm} = \sum_a \lambda_{ia} \braket{\alpha^I_{nlm}}{\chi_a}$ (since all basis are real).
The  density matrix can be represented using  the basis set $\left\{\ket{\alpha^I_{nlm}}\right\}$  as
\begin{equation}\label{eqn:full-D}
    \cD^{II'}_{nlm;n'l'm'} = \sum_i c^I_{inlm} c^{I'}_{in'l'm'}=\sum_i\braket{\alpha^I_{nlm}}{\psi_i}\braket{\psi_i}{\alpha^{I'}_{n'l'm'}}=\sum_{i,a,b}\lambda_{ia}\lambda_{ib}\braket{\alpha^I_{nlm}}{\chi_{a}}\braket{\chi_{b}}{\alpha^{I'}_{n'l'm'}}.
\end{equation}
%As such, we can turn the functional dependence of $E_c$ on $\Gamma(x,x')$, $E_c[\Gamma(x,x')]$, to a function
 
Our task is now reduced to the modeling of $E_c(\{\cD^{II'}_{nlm;n'l'm'}\})$.
%The requirements of locality and symmetry are also transferred  to $E_c(\cD^{II'}_{nlm;n'l'm'})$.
We first discuss the issue of locality and symmetry of $E_c(\cD^{II'}_{nlm;n'l'm'})$.
To our surprise, for all the cases  tested (see next section), it suffices to use the  ``local density matrix''
\begin{equation}
    \left(\cD^I_{nl}\right)_{mm'} = \cD^{II}_{nlm;nlm'} =\sum_i c^I_{inlm} c^I_{inlm'}.
\end{equation}
% \WE{should be improved} \YC{Which one should be improved?}
To deal with  rotational symmetry of the basis $\ket{\alpha^I_{nlm}}$ we use the eigenvalues of the local density matrix
as our descriptor
\begin{equation}
    \label{eqn:d-basic}
    \vd^I_{nl} = \mr{EigenVals}_{mm'} \left[\left(\cD^I_{nl}\right)_{mm'}\right],
\end{equation}
Note that $\vd^I_{nl}$ is a vector with $2l+1$ components. It is also the squares of the singular values of the overlap coefficients
\begin{equation}
    \vd^I_{nl} = \left(\mr{SingularVals}_{im}\left[ c^I_{inlm} \right] \right)^2.
\end{equation}

Finally our model for $E_c$  takes the form:
\begin{equation}
    E_c = \sum_I \epsilon^I_c = \sum_I \cF \left( \vd^I \right)
\end{equation}
where $\vd^I$ denotes the flattened vector made from  $\{\vd^I_{nl}\}$ for all $n$ and $l$. 

%  \WE{we need to say more about the ML part, one subsection}

% For training purposes, the general $\cF$ is separated into two parts, the linear regression term and the neural network correction $\cF = \cF^\mr{Linear} + \cF^\mr{NN}$. 

It is straightforward to verify that the resulting $E_c$ model is symmetry preserving. 
Moreover,  since each atomic contribution 
depends locally on the projection coefficients, the model is, by construction, fairly local and can be trained
using data obtained from small systems.
Finally, all the operations involved here are less expensive than a typical HF calculation.
Therefore, as a tool for post-processing  HF results, the model is very efficient.

In order to approximate the function $\cF$, we propose two widely used machine learning ansatz, linear function $\cF^\mr{lin}$ and neural network $\cF^\mr{nn}$, and determine their parameters by fitting the data.

For linear functions, we have
$$\cF = \cF^\mr{lin} (\vd) = \mb{W} \cdot \vd + \mb{b}$$
where $\mb{W}$ and $\mb{b}$ are parameters learned from data. we use ordinary least square (OLS) regression (or Ridge regression when there are two few training samples) as the training procedure. 

For neural network fitting functions $\cF = \cF^\mr{nn}$, we use standard fully connected feed forward neural network with skip connection\cite{he2016deep} between each layers. The detailed parameters of the neural network model does not influence much the results. In our cases we use 3 hidden layers and 100 neurons per hidden layer. We use GELU \cite{hendrycks2016gaussian} as the activation function for every layer except the final one. The neural network is trained by gradient descent method using Adam optimizer~\cite{kingma2014adam} ($\beta_1=0.9,\,\beta_2=0.999,\,\varepsilon=1\times{10}^{-8}$) with $\ell^2$ loss function and batch size 16. The learning rate is set to $1 \times {10}^{-4}$ at the beginning and decays exponentially with a factor of 0.98 for every 500 epochs. No regularization method is used in training the neural network.

We will see later that for small and simple systems, linear regression is already quite accurate. When
the  systems get larger and more complex, neural network model helps to improve the accuracy. We also note that when the number of training samples is very small,  the neural network model gives worse results, due to overfitting.
% its sample complexity is much higher than linear function. 

{ We remark that unlike conventional post-HF methods, it is not critical to choose HF solutions as the starting point of the DeePHF scheme and to choose the correlation energy as the target.
As we will show with numerical examples, this scheme can also be applied when  DFT is used as the starting point.}

\subsection{Extensions}\label{sec:extdes}
We introduce several extensions of the energy model, which can be used to further improve the 
accuracy of the model. 

First, we may insert a Hermitian operator $\hat{\mathcal{O}}$ in the overlap coefficients
\begin{equation}\label{eqn:c-ext}
    \tilde{c}^I_{inlm} = \bra{\alpha^I_{nlm}}\hat{\mathcal{O}}\ket{\psi_i}.
\end{equation}
As the most straightforward extension, instead of using the bare "local density matrix" $\cD^I_{nl}$, we can 
use a set of kernel functions of the energy $\{f_k(\varepsilon)\}$  by taking $\hat{\mathcal{O}}^2$ to be $f_k(\hat{H}_{HF})$. 
Now the extended ``density matrix'' can be defined as 
\begin{equation}
    \left(\tilde{\cD}^I_{nlk}\right)_{mm'} = \sum_i c^I_{inlm} c^I_{inlm'} f_k(\varepsilon_i).
\end{equation}
There are no specific requirements for the kernel functions $\{f_k\}$ other than that they should be normalized in some way so that shifting the energy will not influence the descriptor. The extended descriptor is defined in the same way:
\begin{equation}\label{eqn:d-ext}
    \tilde{\vd}^I_{nlk} = \mr{EigenVals}_{mm'} \left[\left(\tilde{\cD}^I_{nlk}\right)_{mm'}\right].
\end{equation}
%We will consider $\tilde{\vd}^I_{nlk}$ in our later testing cases.
This will be used later.

Secondly, the choice of the basis set $\braket{\vr_i}{\alpha_{nlm}} = R_n(r_i) Y_{lm} (\theta_i, \phi_i)$,  particularly  $R_n(r_i)$ as well as the operator $\hat{\mathcal{O}}$, can be made adaptive to the training data.
This can be done by introducing some trainable parameters into their expressions.
% to make the feature selection procedure more versatile.
An analogy of such an operation can be found in the construction of the DP model~\cite{zhang2018end}, in which an embedding network is used to define invariant features of the neighboring environment of each atom, and a fitting network is used afterwards to map the features onto the final output.

Finally, more non-local information can be extracted from the density matrix involving two atoms
\begin{equation}
     \left(\cD^{IJ}_{nlk}\right)_{mm'} = w(r^{IJ}) \sum_i c^I_{inlm} c^J_{inlm'} f_k(\varepsilon_i)
\end{equation}
and the descriptor can be calculated accordingly. 
Here $w(r^{IJ})$ is some weight function (of the distance between the two atoms).
To ensure that the eigenvalues stay real we may use instead: 
\begin{equation}
     \cD^{I;\mr{nonlocal}}_{nlk} = \frac{1}{2} \sum_J \left(\cD^{IJ}_{nlk} + \cD^{JI}_{nlk} \right)
\end{equation}
%Yet given the testing cases, the current naive approach would increase generalization gap, so we do not use this construction for now. 

\subsection{An active learning procedure}
Another important problem in our machine learning method is that the labels (in our case the CCSD(T) energies) are very expensive to get. Even a single data point can take hours to calculate. 
Therefore it is important to build the model to be as sample-efficient as possible. To address this issue, we borrow ideas from active learning that aims to learn models with a minimum number of labeled data. 
We need an algorithm that can efficiently go over the unlabeled data and determine which ones should be labeled and put into the training set. 

In order to do this, we train multiple neural network models on the same dataset with different initialization, and use the standard deviation of the output from the different models
\begin{equation}
    \cE = \sqrt{ \left\langle (E_c^\mr{nn} - \left\langle E_c^\mr{nn} \right\rangle)^2 \right\rangle }
\end{equation}
as an indicator for the error in the model. This quantity is called the model deviation in \cite{zhang2019active}. Here $\langle \cdots \rangle$ denotes averaging over the different models. The reason that the model deviation can serve as a good
error indicator is that the landscape for training neural network models is highly non-convex and often highly 
over-parametrized. Therefore different initializations usually lead to different minimizers after the training process. 
For data points close to the training set, the predictions from different minimizers should all be quite accurate
and therefore close to each other, giving rise to small values of the model deviation.  But for those that are far from the the training set, the different minimizers  should give different predictions, resulting in larger values of model 
 deviation. 

% The model deviation $\cE$ can be easily calculated on any unlabeled data. 
The idea is to simply choose  data points with the largest value of model deviation to be
 labeled and add them to the training set. 
%This process can be repeated, selecting out only the points that our model is not predicting well. 
This allows us to work with the most representative subsamples of the unlabeled dataset and thereby improving the
sample-efficiency.
% relatively small amount of data, achieving higher sample efficiency.

\subsection{Comparison with state-of-the-art methods}
%Here we would like to compare our method with other reported methods following the similar approach. 
The methodology presented here is an alternative version of MOB-ML, although our ultimate goal might
be different from that of MOB-ML.
%MOB-ML is the first method we know to fit correlation energy using orbitals from HF solution. It exhibits good accuracy and transferability on a relatively large dataset.
MOB-ML uses features that come from the matrix elements of Fock, Coulomb and exchange operators on a set of localized molecular orbitals. Both occupied and virtual orbitals are involved. 
In DeePHF, the features come directly from the projected density matrix of the ground state wave function.
No gauge transformation for localized molecular orbitals is used. 
%The handling of labels is also different. 
Secondly, in MOB-ML, the correlation energy is decomposed into  contribution from each orbitals and the per orbital value is directly used as the label. 
In DeePHF, we only use correlation energy itself as the label. 
Finally, instead of using kernel based methods such as the Gaussian process regression, we use linear regression or neural network model as the fitting function.

There are also methods aiming at parameterizing an exchange-correlation functional in Kohn-Sham DFT approach. 
NeuralXC~\cite{dick2020machine} defines such a functional by neural network which takes as input the projected density onto an atom-centered basis set, and employs a self-consistent-field (SCF) procedure to train it. DeePHF, on the other hand, uses projected density matrix that contains more information. We also examined the DeePHF method on a large dataset containing thousands of different molecules, which is not shown in the NeuralXC article~\cite{dick2020machine}.

\section{Results \& Discussion}
To examine the performance of the DeePHF scheme, we follow the testing procedure in~\cite{cheng2019universal}, utilizing their publicly available dataset~\cite{cheng2019data}. We also included a larger dataset ANI-1ccx\cite{smith2019approaching,smith2020ani} to further test the transferability of the model. We train and test our model respectively on the following sets of data: (i) single water molecules, (ii) short alkanes, and (iii) small organic molecules. All training and testing data are disjoint. Unless otherwise specified, we use the CCSD(T) correlation energies  as labels. In order to obtain the descriptors, we use the geometries provided by the dataset to perform the Hartree-Fock calculation. The calculation is done using PySCF \cite{sun2018pyscf} with cc-pVTZ basis \cite{woon1993gaussian}. The results for  HF energies are very close 
to the values provided in the dataset -- the differences are normally smaller than 0.1~mH and should be a result of numerical error. This difference should have no influence on our results. %The GDB-13-T data in (iii) is not used since it does not contain CCSD(T) energies.

In order to speed up the calculation, the radial part of the  basis function  is constructed by linear combinations of Gaussians centered at zero, compatible with the commonly used Gaussian-type orbitals (GTOs) in quantum chemistry. The angular index $l$ is taken to be $l \in \{0, 1, 2\}$, resulting in 1, 3 or 5  values of $m$
respectively. Unlike the traditional GTOs, we use the same set of radial functions for all angular indices. The detailed coefficients can be found in the Appendix~\ref{appendix:basis}. Normally we use 12 different radial functions, resulting in a total of $12 \times (1 + 3 + 5) = 108$  basis functions per atom. However for the test on short alkanes, due to the limited number of training data, we use 9 radial functions thus 81  basis functions. It is worth noting that as long as 
a sufficient number of radial functions are  used, the specific number of Gaussians and the associated parameters have little influence on the results. 
%The reason to use 108 and 81 is just that the authors feel that they are adorable numbers.

\subsection{Simple Systems}
We first examine the performance of DeePHF for relatively simple systems, namely (i) single water molecules and (ii) short alkanes. Since these systems are simple enough, we only use our basic descriptor $ \{ \vd_{nl}^I \} $ defined in Eq~\ref{eqn:d-basic} and use a linear regression $\cF = \cF^\mr{lin} $ to fit the correlation energy. Extended descriptor and more complex fitting functions can be used, but they 
do not give significantly better  test accuracy.

The first system we tested is single water molecules.  We trained our model on different numbers of randomly drawn samples from 1000 water geometries in the dataset, and used the rest  as the testing set.  We examined the  mean absolute error (MAE) and max absolute error (MaxAE) along the learning curve. When the training set is smaller than the dimension of the descriptors (108), we added a small ridge regularization coefficient ($\alpha = 10^{-9}$). We repeated these steps for 1000 times and report the average MAE.

\begin{figure}[htbp]
    \centering
    \includegraphics[width=0.6\textwidth]{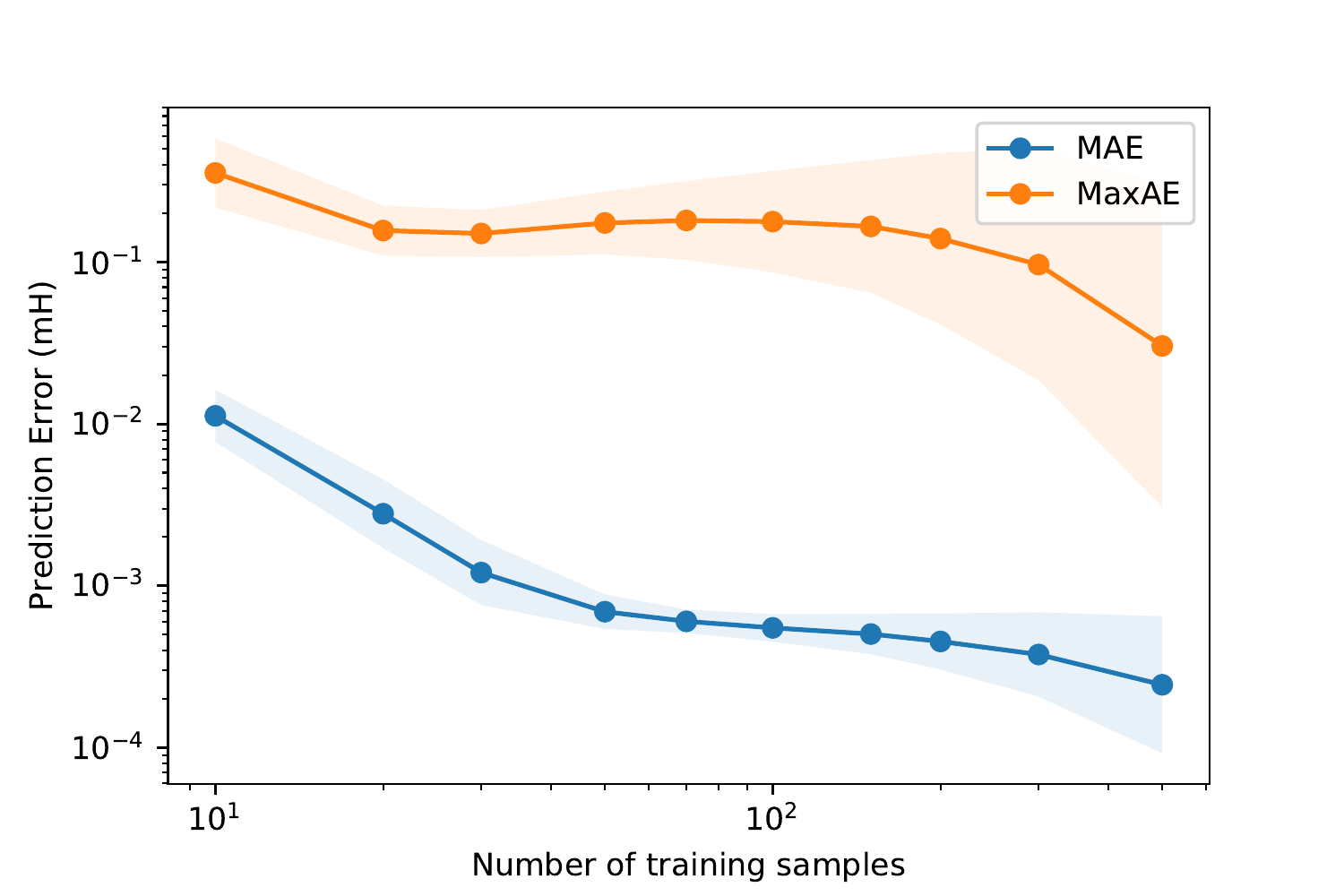}
    \caption{The learning curve of DeePHF for water molecules. Both mean absolute error (MAE) and max absolute error are shown (MaxAE). Shaded area shows the size of the standard deviation of 1000 different runs. }
    \label{fig:waterlc}
\end{figure}

The resulted learning curve is shown in Fig.~\ref{fig:waterlc}. Generally speaking, the results are comparable with that of  MOB-ML ~\cite{cheng2019universal}. 
With only 10 training samples we can predict the energy of all the testing data  with errors less than 1 mH. 
Adding more training samples will result in smaller mean absolute error.
However the max absolute error behaves differently and has a much larger variance. 
This is because the max absolute error is strongly influenced by the few specific data points whose errors may be hard to reduce by adding training data.

%  \WE{what are those outliers? why are they outliers?}

Another relatively simple dataset we examined is short alkanes. The dataset contains 100 geometries of methane, 1000 of ethane, 1001 of propane and 101 of both n-butane and isobutane. This dataset
has been used to test the transferability of the model by many authors, e.g, Refs.~\citenum{cheng2019universal} and \citenum{dick2020machine}. 
%However, as we will show below, it is not a good standard at least in the case of our model. 

The same linear fitting function and basic descriptor was used as for the case of water. Following the procedure in \cite{cheng2019universal}, we randomly chose 50 ethane molecules and 20 propane molecules to train a linear model by OLS  with a small ridge regularization coefficient $\alpha = 10^{-7}$ and normalized descriptors. 
We then used this model to predict the correlation energy of n-butane and isobutane molecules. 
We repeated the training process for 10000 times and recorded the resulted mean error (ME), mean absolute error (MAE), max absolute error (MaxAE) and mean absolute error after applying a global shift for each type of molecule that sets the mean error to zero (relative error, MARE).

\begin{figure}[htbp]
    \centering
    \begin{subfigure}{0.45\textwidth}
    \includegraphics[width=\textwidth]{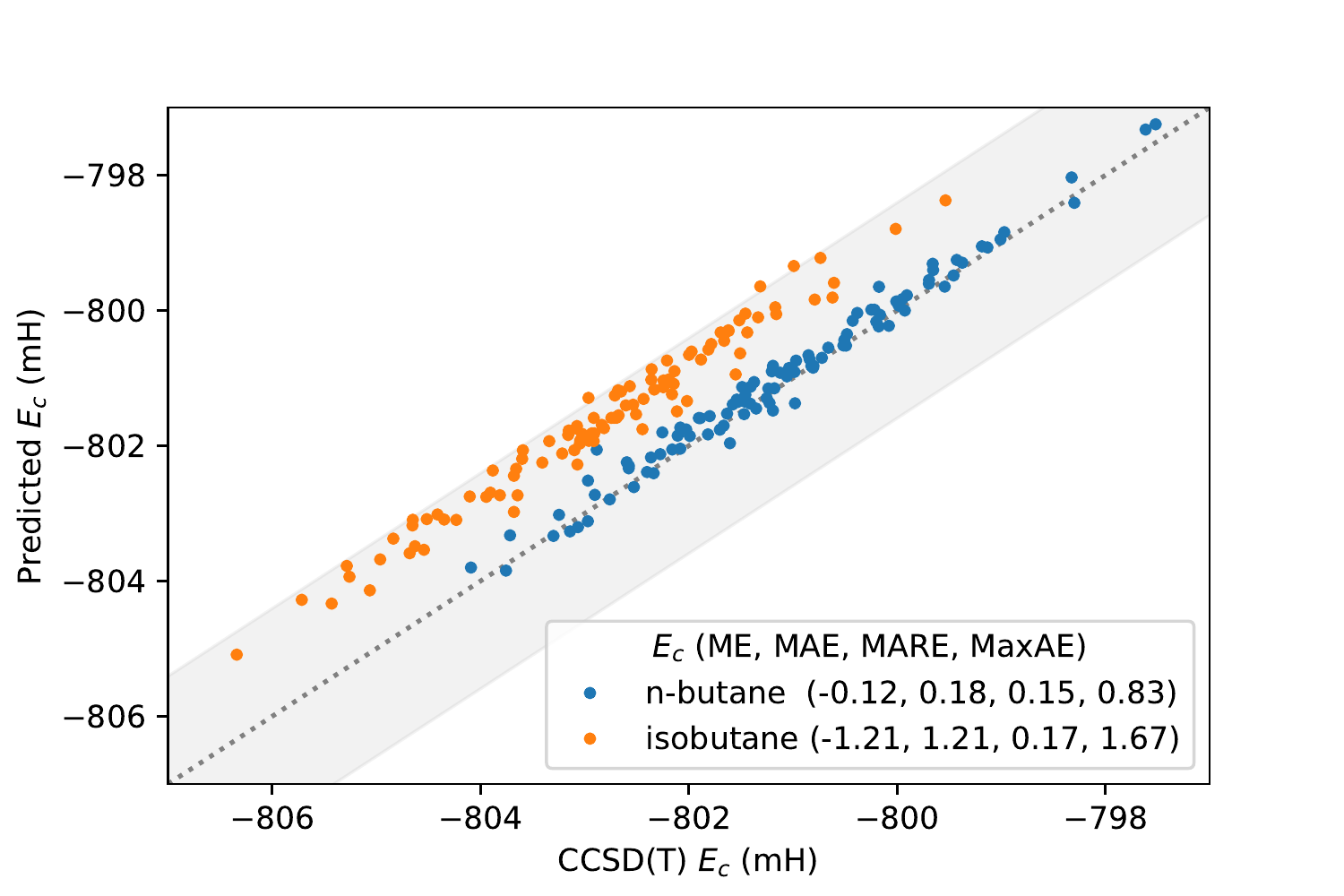}
    \end{subfigure}
    \begin{subfigure}{0.45\textwidth}
    \includegraphics[width=\textwidth]{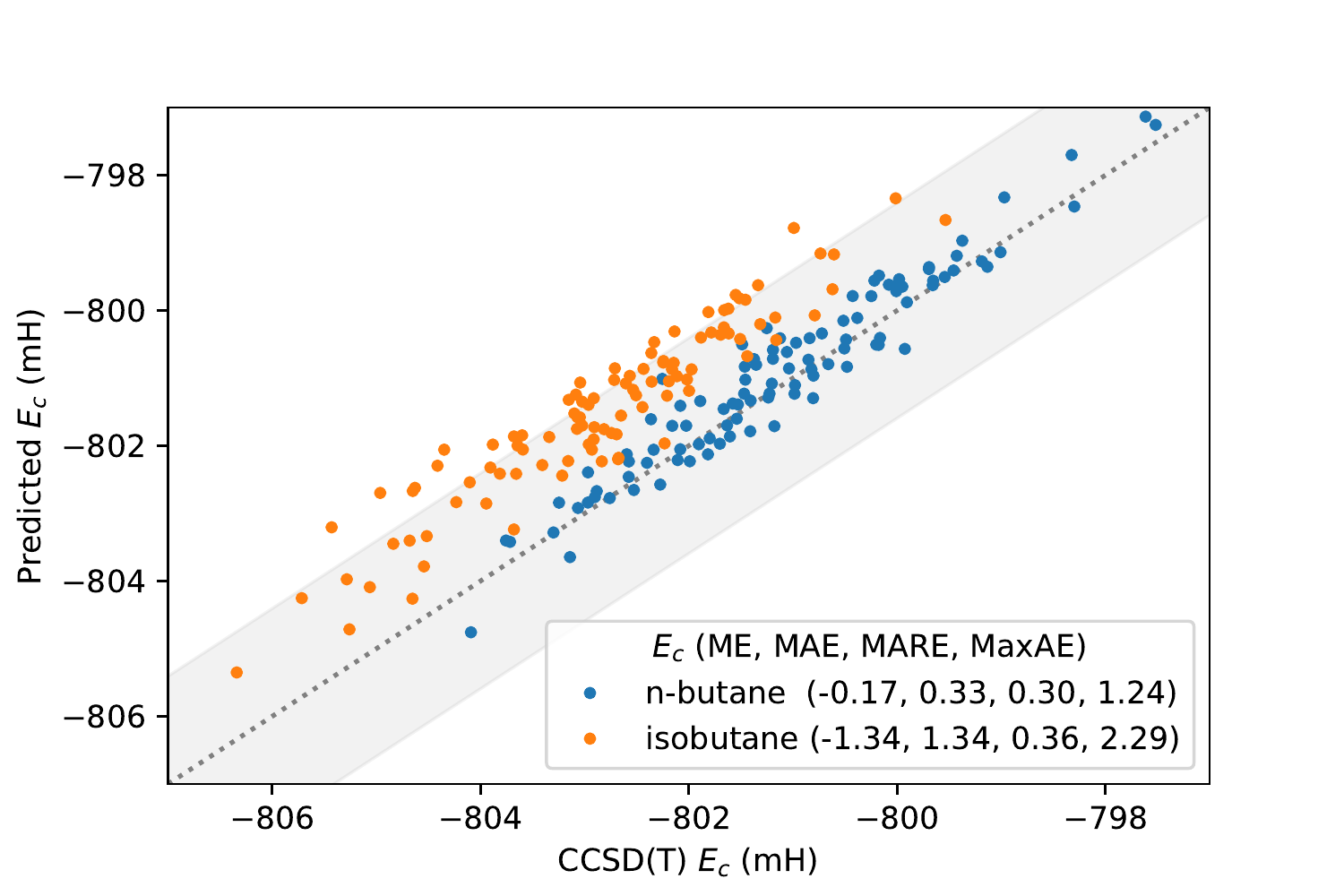}
    \end{subfigure}
    \caption{Predicted correlation energy of DeePHF versus the true CCSD(T) energy of n-butane and isobutane. Two different results from different training samples are shown. The one on the left is a relatively better result.
    The one on the right is worse.  But these are just  results of randomly sampled training datasets.
    Grey shaded area corresponds to the region where the error is smaller than chemical accuracy (1 kcal/mol). 
    No global shift is applied on these values.
 %    Statistical values including mean error (ME), mean absolute error (MAE), mean absolute error after global shift (MARE) and max absolute error (MaxAE) are presented on legend.
 }
    \label{fig:alkaneec}
\end{figure}

\begin{figure}[htbp]
    \centering
    \includegraphics[width=\textwidth]{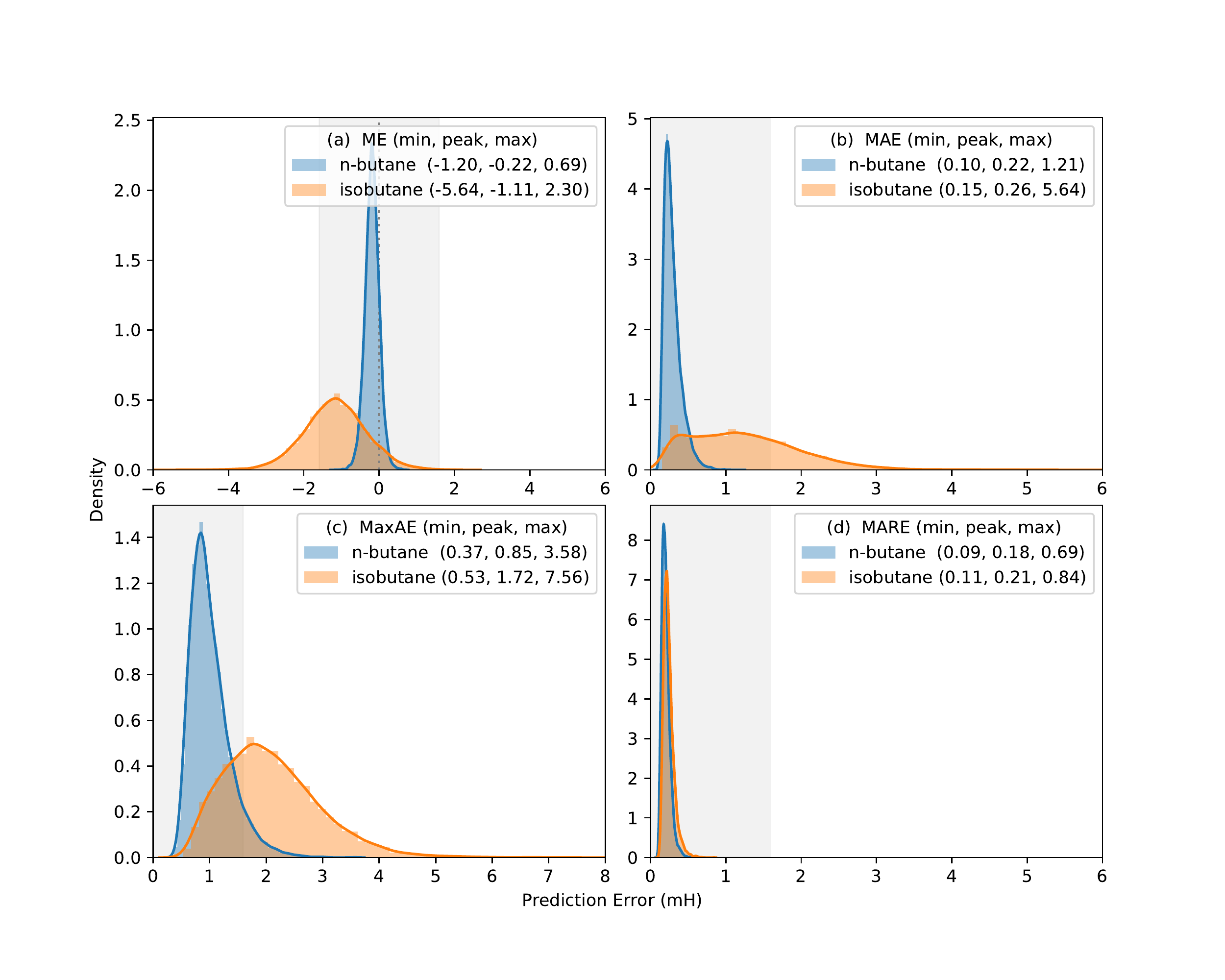}
    \caption{Distributions of prediction error based on different metrics when DeePHF is tested on n-butane and isobutane, including (a) mean error (ME), (b) mean absolute error (MAE), (c) max absolute error (MaxAE) and (d) mean absolute error after global shift (MARE). Grey shaded area corresponds to the region where the error value is smaller than chemical accuracy (1 kcal/mol). 
    The minimum, the peak and the maximum values of the distribution are shown in the legend. Note the ranges of X axis are different, except for (b) and (d).}
    \label{fig:alkanedist}
\end{figure}

The distributions of these errors is shown in Fig.~\ref{fig:alkanedist}.
If we only look at the error statistics in the better one of Fig.~\ref{fig:alkaneec} or the peak values of the distributions in Fig.~\ref{fig:alkanedist}, we might conclude that the model transfers well: almost all the predictions are within chemical accuracy to the ground truth. These results also outperform the current best results reported in Refs.~\citenum{cheng2019universal} and \citenum{dick2020machine} (Since  relative energies were used in these literature, we have to  use the MARE metric in order to
compare with their results). A detailed comparison 
can be found in Table~\ref{tab:butane}.

\begin{table}[htbp]
\centering
\begin{tabular}{l||l|l}
Methods     & N-butane & Isobutane \\
\hline
MOB-ML\cite{cheng2019universal}      & 0.32     & 0.33      \\
NeuralXC\cite{dick2020machine}    & 0.24     & 0.22      \\
DeePHF (peak) & 0.18     & 0.21      \\
DeePHF (best) & 0.09     & 0.11     
\end{tabular}
\caption{Mean absolute error after global shift (MARE) for small alkanes system. For each model,  50 randomly sampled  ethane and 20 propane
configurations are used  the training set and was tested on n-butane and isobutane. Global shift has been applied for each molecular type 
so that the mean prediction error is 0. Both the most probable value (peak) and the best result are shown for the 
proposed model, based on different randomly sampled training sets. Errors are given in mH.}
\label{tab:butane}
\end{table}

However,  the other results shown in Fig.~\ref{fig:alkaneec} are much worse, suggesting that there is a large variance in the test error for these models
when different random samples of the training data are selected. This also suggests that the transferability concluded above is not really genuine.

%Yet the only difference between these two results is that they are trained on two different training sets, both randomly sampled as described above. 
After a closer look we notice that both Fig.~\ref{fig:alkanedist}(a) and Fig.~\ref{fig:alkaneec} exhibit constant bias on the predicted energies.  
%Also, without applying a molecular specific correction for that bias to zero the mean error, the error statistics in Fig~\ref{fig:alkanedist}(a)(b)(c) form a rather wide distribution, meaning that these error metrics heavily depend on the sample we pick in the training set. 
This problem is much more severe for the case of isobutane. 
One can apply a sample-dependent global shift to reduce the bias, as was done in Refs.~\citenum{cheng2019universal} and \citenum{dick2020machine}.
Indeed we also found that 
%On the other hand, although
 the  relative error  with  such a global shift applied has much better distribution.
However, this is not really practical since the amount of the shift required depends information from the testing set.
%In order to know the amount of the shift , we must know the labels of the testing set.
%  \WE{???} 

One intuitive explanation for the error distribution is to consider the atomic environment from which we build our descriptor. For isobutane, there is a carbon atom connecting with three other carbons. Such environment does not occur in neither ethane nor propane. Hence the model have no clue how much this atom should contribute to the correlation energy. In other words, this is the extrapolation case, so there's usually a systematic error in the predicted energy (the large global shift in isobutane) and the prediction depends largely on how the training data is selected (the wide error distribution). On the other hand, n-butane does not contains such an atom. All its atoms have similar ones in ethane and propane and the prediction is more like interpolation therefore behaves much better. 

This explanation can be partially justified by the observation that the test accuracy was actually  \emph{reduced} when more training samples from methane are included. The carbon atom in methane connects to four hydrogen atoms and does not appear in all other molecules. When the amount of training data is very limited and concentrated on a specific type of molecules, adding such data would actually introduce some bias in the model and worsen the results, especially in linear model.

\begin{figure}[htbp]
    \centering
    \includegraphics[width=0.6\textwidth]{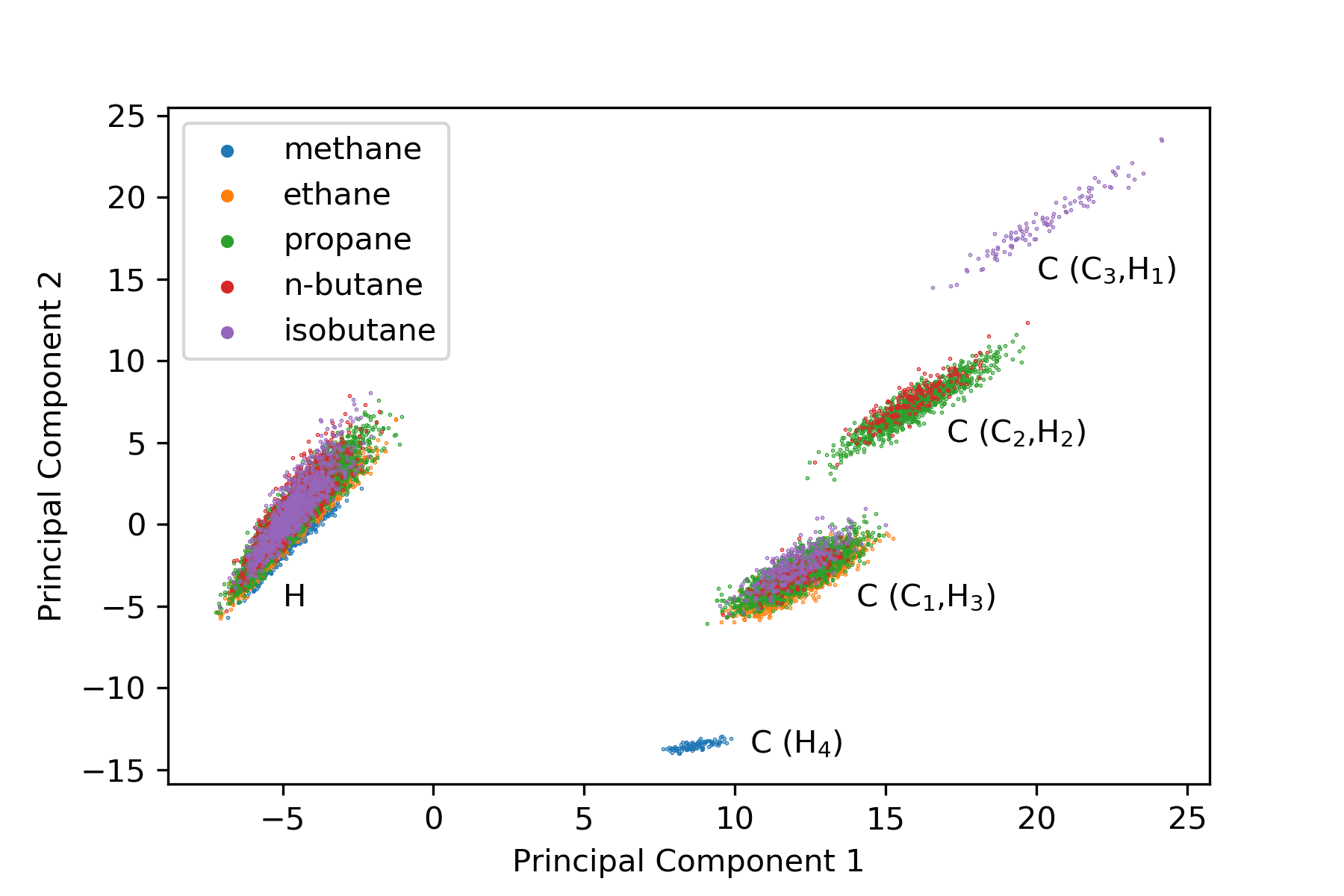}
    \caption{Projected descriptors of small alkane molecules. Each dot on the graph stands for a single atom, with certain color to identify the molecule it belongs to. Atom type is annotated near the corresponding cluster, including neighboring environments for carbon atoms. }
    \label{fig:alkanepca}
\end{figure}

To see it more clearly, we project the descriptors into two dimension using principal component analysis (PCA) and plot them in Fig.~\ref{fig:alkanepca}. Atoms separate clearly into different clusters based on their element type and neighboring atoms. We can see that our training set (ethane and propane) covers the hydrogen cluster and two carbon clusters ((C${}_1$,H${}_3$) and (C${}_2$,H${}_2$)), as expected. The atoms from n-butane lies just in these three clusters hence the interpolation regime of our model, while some atoms from isobutane form another carbon cluster (C${}_3$,H${}_1$) that is outside our training set, corresponding the extrapolation regime. Moreover, the carbon atoms from methane form yet another cluster that is away from both our training and testing set. This behavior is exactly what we expect from our previous explanation.
Therefore, we believe that the apparent "transferability"  found above
is basically due to data selection and is not an indication of real  transferability for larger systems.

\subsection{The QM7b-T dataset}

To test the transferability of DeePHF in a more realistic setting, we utilized the QM7b-T dataset.  
%We  focus on the CCSD(T) energies and exclude the GDB-13-T dataset  since we do not have CCSD(T) energies for them.
 Again, we randomly select different numbers of  samples to train the model and test its accuracy on the rest of the dataset. 

In this dataset, the system is complex enough that linear regression is no longer adequate.
%we would  reach the limit of expressive power of linear fitting function $\cF^\mr{lin}$. 
Therefore, we introduce a neural network model as an improved fitting function $\cF = \cF^\mr{nn}$. 
Since  the number of training samples is large enough, we were able to test the extended descriptor $ \{ \tilde{\vd}_{nl}^I \} $ described in Eq.~\ref{eqn:d-ext}. We used two simple kernel functions to calculate the extended density matrix, namely identity function $f_1(\varepsilon) = \varepsilon$ and the exponential function $f_2(\varepsilon) = e^\varepsilon$. Combined with the basic descriptors which can be viewed as using constant kernel $f_0(\varepsilon) = 1$, the total number of descriptors is three times larger than the case of basic descriptors.  %\WE{than what?}

\begin{figure}[htbp]
    \centering
    % \begin{subfigure}{0.45\textwidth}
    % \includegraphics[width=\textwidth]{img/sample_curve.pdf}
    % \end{subfigure}
    % \begin{subfigure}{0.45\textwidth}
    \includegraphics[width=0.6\textwidth]{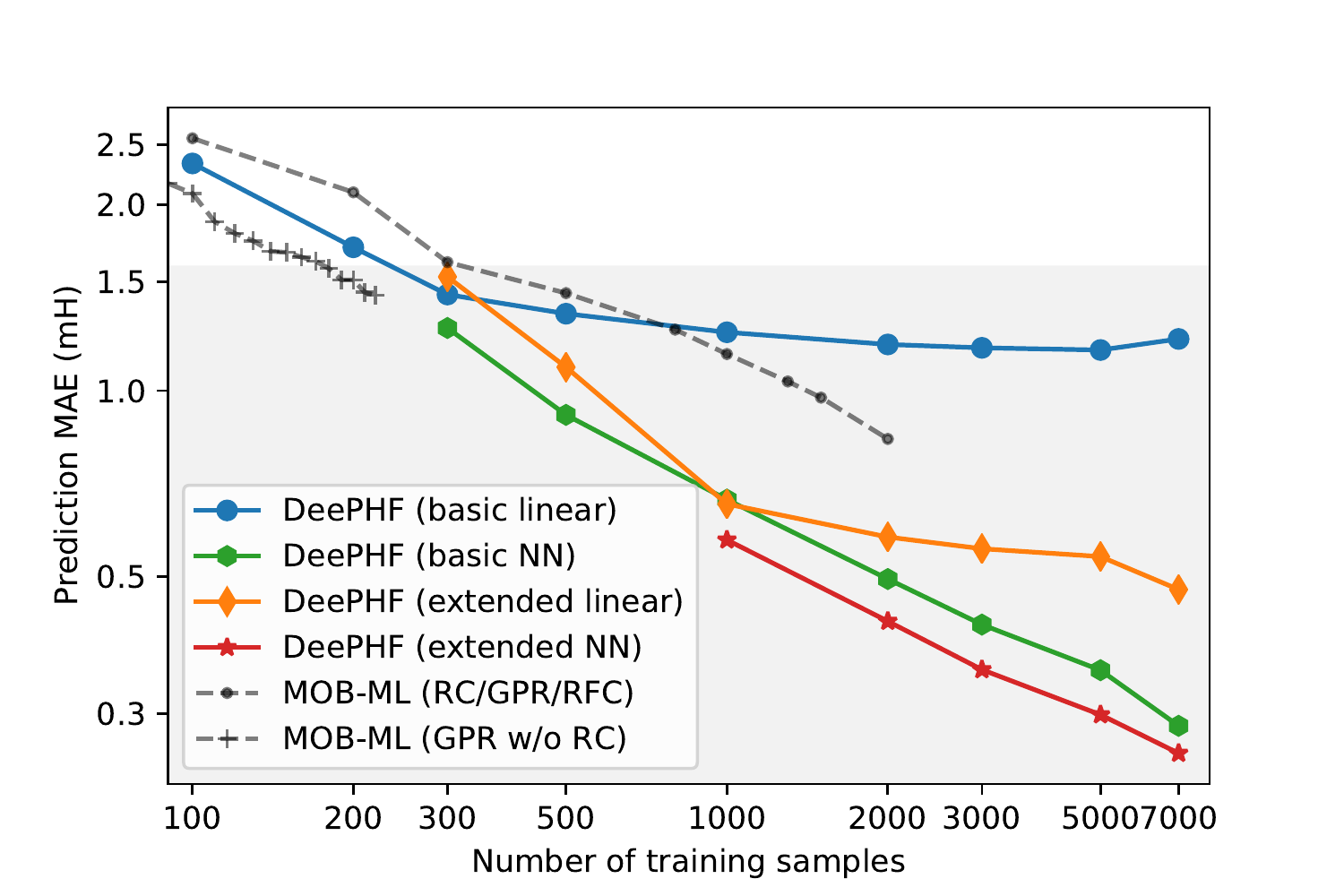}
    % \end{subfigure}
    \caption{
    The learning curve of DeePHF on the QM7b-T dataset. Depending on whether the extended descriptor and  neural network fitting function is used, results for four different constructions are presented. Results from Ref.~\citenum{cheng2019regression} are also included for comparison. Grey shaded area corresponds to the region where the error is smaller than chemical accuracy (1 kcal/mol). 
    % (left) Linear scale. (right) Log scale. 
    }
    \label{fig:qm7lc}
\end{figure}

\begin{figure}[htbp]
    \centering
    \includegraphics[width=\textwidth]{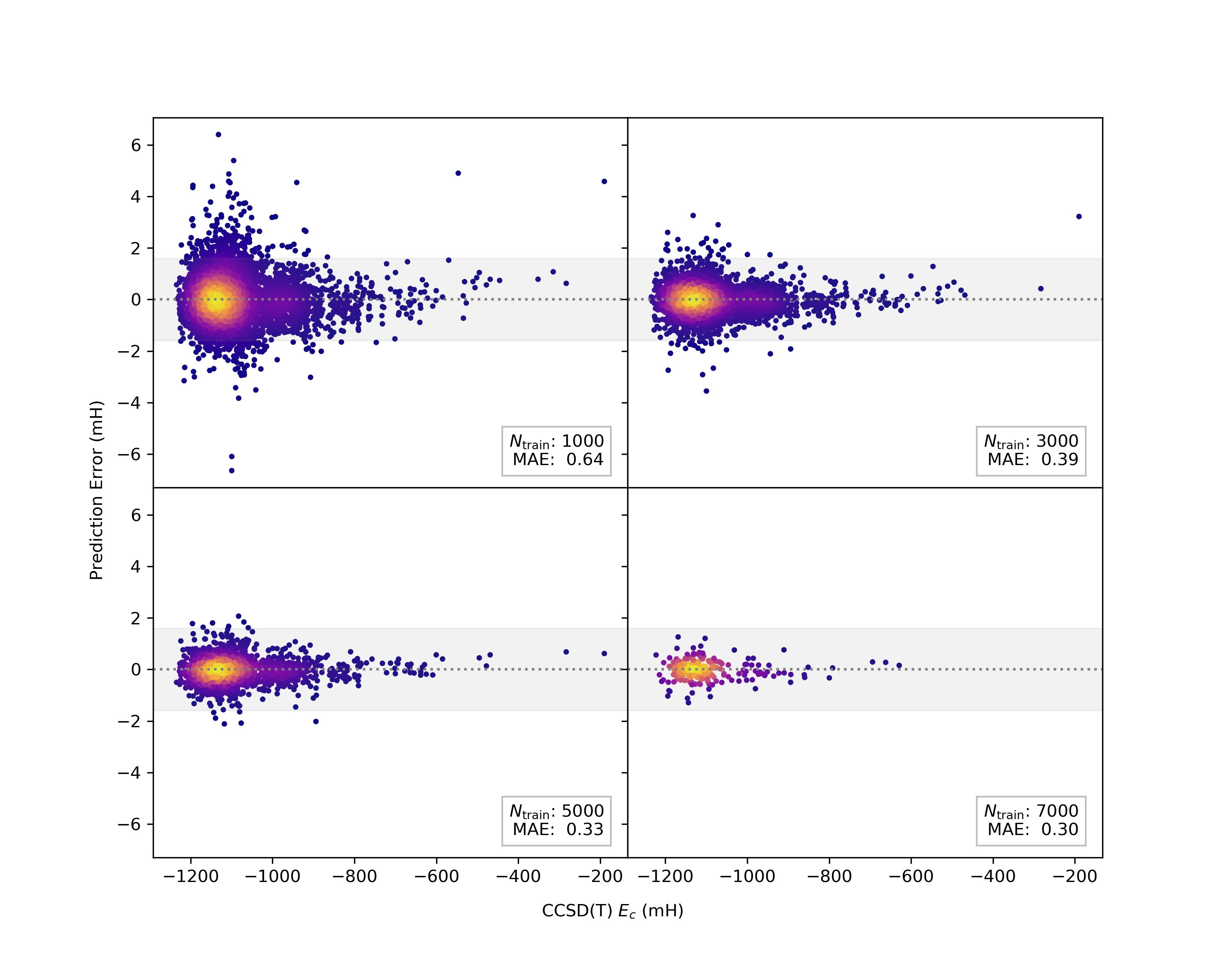}
    \caption{Testing errors of all testing molecules for DeePHF trained on different number of training data. The model is built with extended descriptors and uses neural network as the fitting function. 
    Since we always use the remainder of the dataset as the testing set,  the number of testing data  decreases when we use more training data. Grey shaded area corresponds to the region where the error is smaller than chemical accuracy (1 kcal/mol). Color on the points represents the density of the points.}
    \label{fig:qm7ec}
\end{figure}

With basic or extended descriptors and linear or neural network fitting functions, we have a total of four scenarios. We examine the learning curve for all the four scenarios and the results are shown in Fig~\ref{fig:qm7lc}. 
The linear model is only used when the number of samples is roughly no less than the number of descriptors,
and the neural network model is used when the number of samples is three times larger. For training samples less than these levels, we should not expect the model to perform well. 
Due to the cost of training neural network models, we do not test the effect of different selections of the training data,
as we did for the case  of water. Except for the curve reported in Ref.~\citenum{cheng2019regression}, all models are trained on the same set of data.  When the dataset is augmented,  existing samples in the dataset are kept.

Not surprisingly, given sufficiently many training samples, the neural net model with extended descriptor performs 
the best. The performance of the linear model saturates quickly
when the number of training samples increases, indicating the need of using more sophisticated models. Neural network models, on the other hand, exhibit the expected power-law behavior\cite{hestness2017deep}, suggesting that  further improvement of the accuracy is possible by adding more data. At 7000 training samples we notice that the  model predictions are less consistent, likely due to the small size of the testing set. 

We also compare our results with MOB-ML method. Two versions of MOB-ML are included, namely the one with regression clustering, Gaussian process regression and random forest classification (RC/GPR/RFC), and the one without clustering (GPR without RC). RC/GPR/RFC version reaches the best testing accuracy regardless of number of training samples while GPR without RC version performs better in terms of sample efficiency.
In general, our model is comparable with MOB-ML method. To be more specific, our model outperforms RC/GPR/RFC version in all cases as long as we use scenarios that are expressive enough. On the other hand, DeePHF performs slightly worse than GPR without RC version when the size of training set is small.
One possible reason is that MOB-ML uses per-orbital pair correlation energies as training labels, which can be dozens of times more than a single correlation energy value. This additional information from label can be very important when training data is relatively sparse.

To get a more intuitive view, we summarize in Fig.~\ref{fig:qm7ec} all the testing results using the neural network model trained on different number of samples. We  see that when using only 1000 samples, although the mean absolute energy is 0.64, better than any other model reported, we still get a lot of points with pretty large individual error. When we increase the number of training samples, the testing error decreases as expected, so is the number of points that lie outside the chemical accuracy region. When the size of the training set reaches 5000, the results are very close to chemical accuracy. 
When the training set increases to 7000  samples,  all testing results lie within the chemical accuracy region. 

To explore the sample efficiency of our model, we examine the active learning procedure on the QM7b-T dataset, using the best performing scenario: neural network fitting function with extended descriptor. 
At this point, this exercise serves as a proof of concept. We do not calculate new CCSD(T) data. Instead, we iteratively select and add existing data into training set based on the error indicator calculated using model deviation. The detailed steps are as follows: First we use the same 1000 training samples to train four models  independently. Next we predict the energy using all four models on the remaining data in QM7b-T. We  select another 200 molecules with the largest
values of  model deviation, adding them to the training set and repeat the same procedure. 
After 20 rounds we reach a training set of 5000 samples.

\begin{figure}[htbp]
    \centering
    \includegraphics[width=0.6\textwidth]{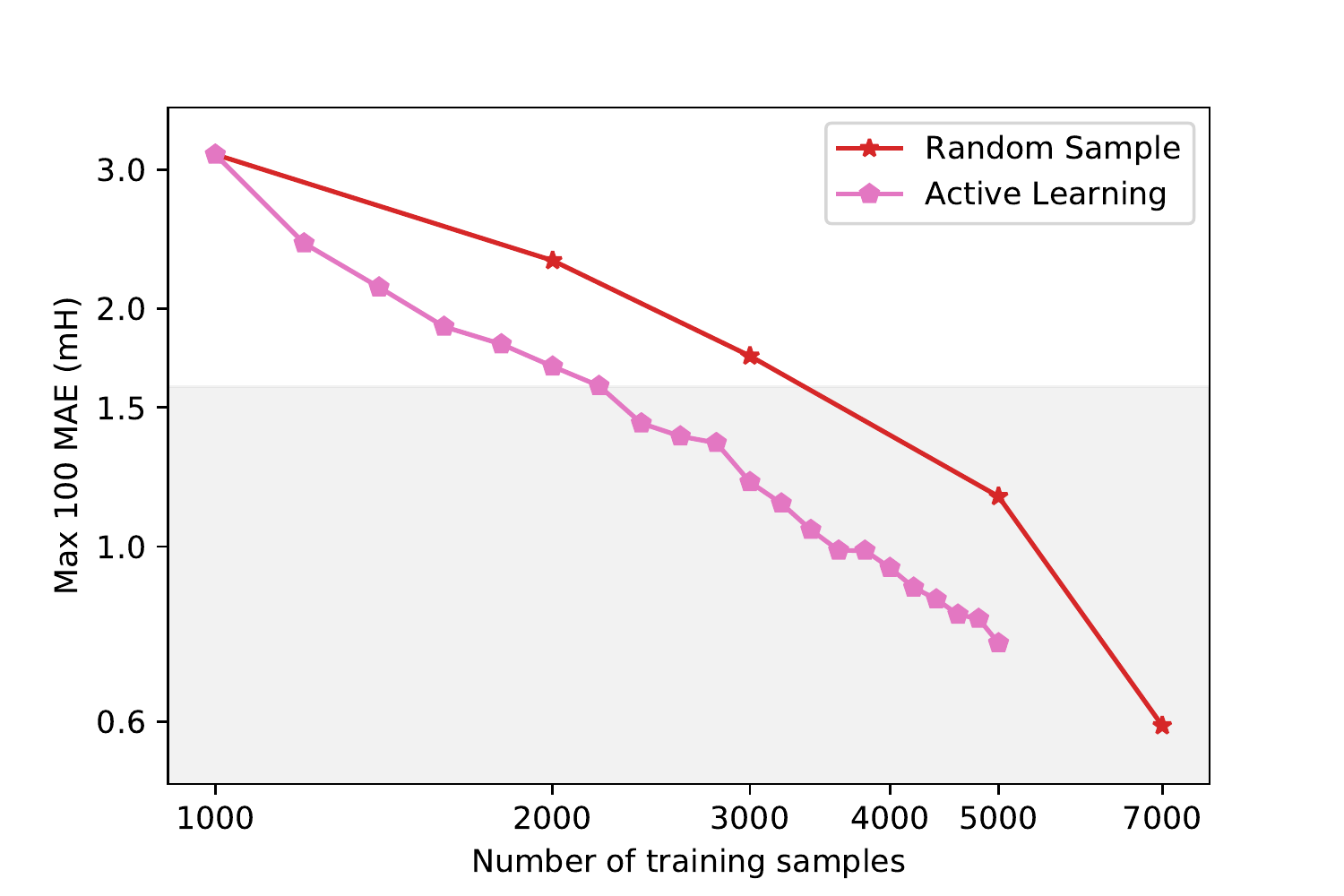}
    \caption{The learning curve of DeePHF with active learning on the QM7b-T dataset. Extended descriptor with neural network fitting function is used. The curve for randomly selected samples is same as the one shown in Fig~\ref{fig:qm7lc} and is used here for comparison. Grey shaded area indicates the region where the error is smaller than chemical accuracy (1 kcal/mol). 
    }
    \label{fig:qm7allc}
\end{figure}

\begin{figure}[htbp]
    \centering
    \begin{subfigure}{\textwidth}
    \includegraphics[width=\textwidth]{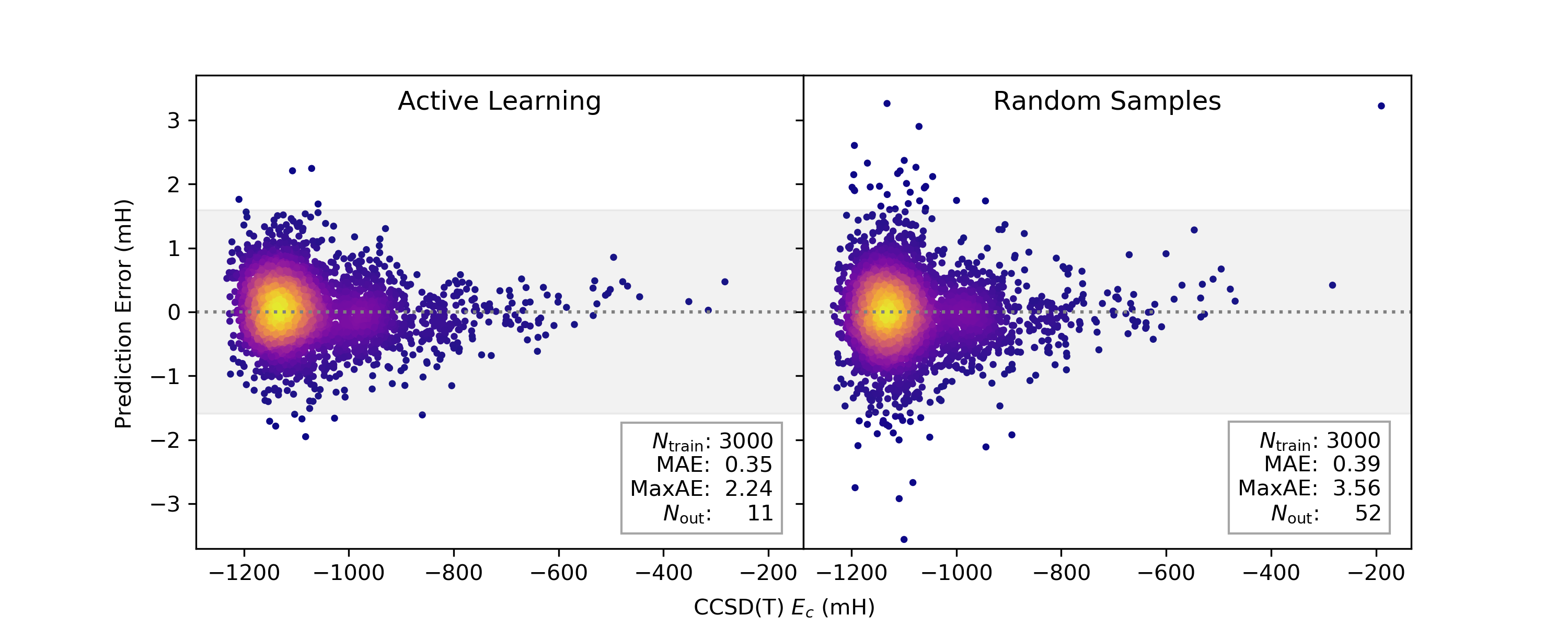}
    \end{subfigure}
    \begin{subfigure}{\textwidth}
    \includegraphics[width=\textwidth]{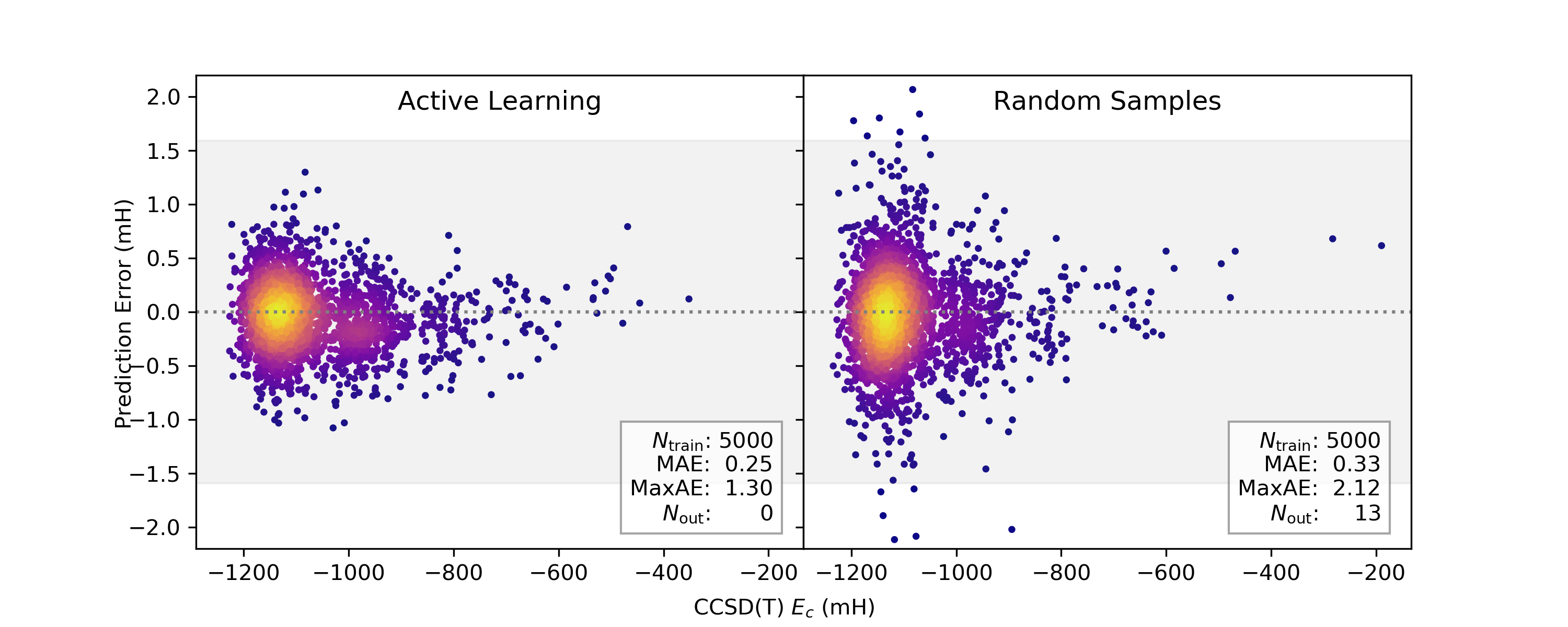}
    \end{subfigure}
    \caption{Scatter plots of the error for the testing samples with and without active learning. The size of the training set is 3000 and 5000, respectively, for the upper and lower figure. Error statistics include  the mean absolute error (MAE), max absolute error (MaxAE) and the number of points with error larger than chemical accuracy (1 kcal/mol) ($N_\mr{out}$). Grey shaded area indicates the region where the error is smaller than chemical accuracy. Color on points represents to the density of the points.}
    \label{fig:activelearn}
\end{figure}

The resulted learning curve is shown  in Fig~\ref{fig:qm7allc}. A comparison of the testing error between actively learned model and model trained on randomly sampled data is also shown in Fig~\ref{fig:activelearn}. 
For the model with active learning, we used the mean absolute error of the largest 100 points over the entire dataset (Max 100 MAE) as our error metric. We  see that the error is significantly reduced by active learning on the whole learning curve. The number of training samples needed to reach chemical accuracy is also  reduced. Note that
 in the case of 7000 training samples, the number of data outside the training set is only 211. Hence the Max 100 MAE exhibits a large reduction due to the lack of data. This is not expected to happen if we  used larger dataset.

A more concrete illustration can be found in Fig~\ref{fig:activelearn}. Active learning also largely reduces the max absolute error and the number of samples that lie  outside the chemical accuracy region. 
This points to the possibility of using active learning  to continuously improve our model by adaptively choosing and labeling new samples on a much larger unlabeled dataset. We intend to pursue this in the future and we
invite interested readers to collaborate on this project.

To further examine the transferability of our DeePHF model, we also test the (QM7b-T trained) model on the GDB-13-T dataset. We follow a similar approach in Refs.~\citenum{cheng2019universal} and \citenum{cheng2019regression} by training the model on randomly sampled data from QM7b-T and testing the model accuracy on GDB-13-T. We use the same set of data as that in the study of the learning curve (Fig.~\ref{fig:qm7lc}), only that the training and testing labels are MP2 correlation energy instead of CCSD(T), since we do not have CCSD(T) labels for the GDB-13-T dataset. We 
examine two scenarios, neural network model with basic or extended descriptor. No active learning procedure is used in this test.

\begin{figure}[htbp]
    \centering
    \includegraphics[width=0.6\textwidth]{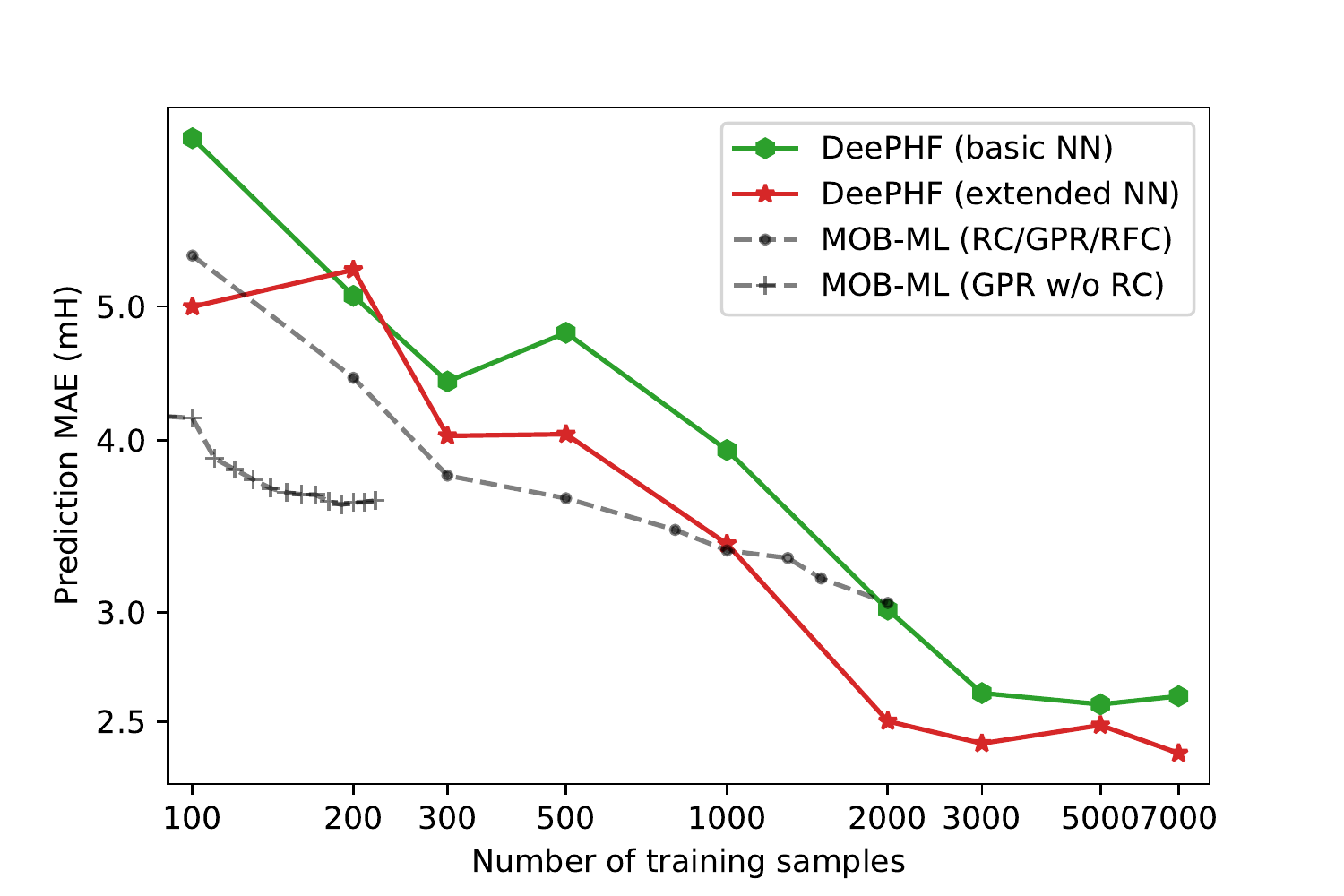}
    \caption{
    The learning curve of DeePHF by training on the QM7b-T dataset and testing on GDB-13-T dataset. The extended descriptor and  neural network fitting function is used. Results from Ref.~\citenum{cheng2019regression} are also included for comparison.}
    \label{fig:gdb13lc}
\end{figure}

The results can be found in Fig.~\ref{fig:gdb13lc}. In general it is comparable with that of the MOB-ML (RC/GPR/RFC) method. With a sufficiently large training set, our model performs relatively better. Using extended descriptor outperforms MOB-ML (RC/GPR/RFC) in large training set regime. When the number of  the training samples is less than 1000, the error of DeePHF is relatively large, possibly due to the sparsity of  the data. 

On the other hand, in all cases, the errors are larger than  chemical accuracy. The anomalous zig-zag behavior in the learning curve (the \emph{decrease} of accuracy when adding more training data) implies that the testing performance is largely decided by the selection of data. Furthermore, the testing error exhibits a large constant shift ($\sim 1.5$mH) on the whole GDB-13-T dataset. This behavior indicates that the test on the GDB-13-T dataset is likely in the extrapolation regime. There should be some information not captured by the training set (QM7b-T). In other words, the data from QM7b-T is insufficient for getting a robust model  on the GDB-13-T dataset that contains larger molecules.

\begin{figure}[htbp]
    \centering
    \includegraphics[width=0.6\textwidth]{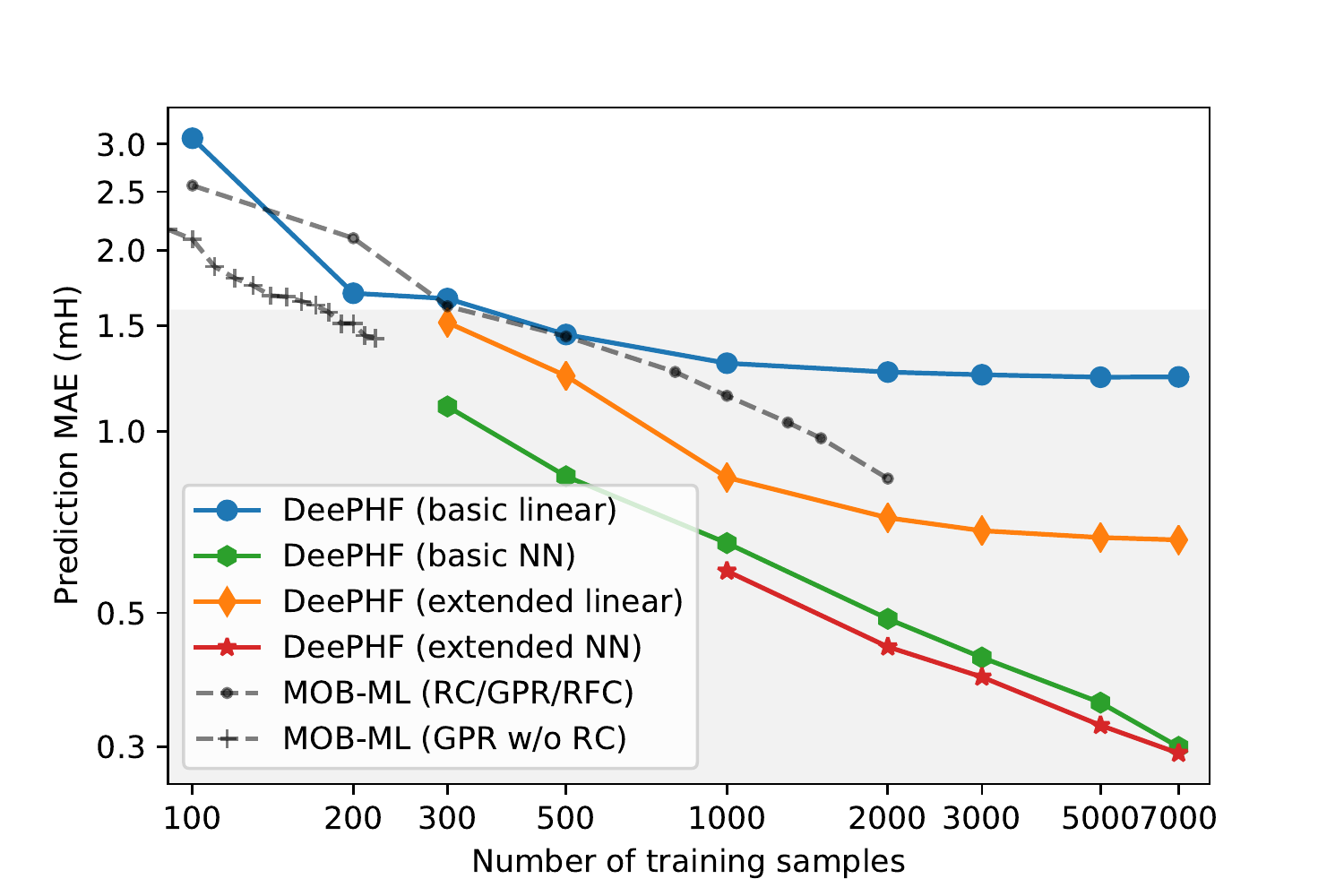}
    \caption{
    The learning curve of the DeePHF method on QM7b-T dataset using DFT  with PBE functional as 
    the starting point. Depending on whether the extended descriptors and the neural network
    model fitting functions are used, results from four different constructions are presented. Results from Ref.~\citenum{cheng2019regression} are also included for comparison. Grey shaded area indicates the region where the error is smaller than chemical accuracy (1 kcal/mol).
    }
    \label{fig:pbeqm7lc}
\end{figure}

{As we have mentioned earlier}, the proposed scheme is not limited to starting with the HF solution and fitting the  correlation energy $E_c = E_\mr{CCSD(T)} - E_\mr{HF}$. 
Other self consistent models such as KS-DFT can also be used as the starting point and 
similar ``DeePHF'' models can be trained to predict the modified ``correlation'' energy 
$E'_c = E_\mr{CCSD(T)} - E_\mr{KS}$. 
Here we demonstrate an example of using KS-DFT with the PBE functional as the starting point. The resulted learning curve can be found in Fig.~\ref{fig:pbeqm7lc}. Although the linear fitting results are not as good, the neural network results are comparable and even slightly better than that of the HF model.

\begin{figure}[htbp]
    \centering
    \includegraphics[width=0.6\textwidth]{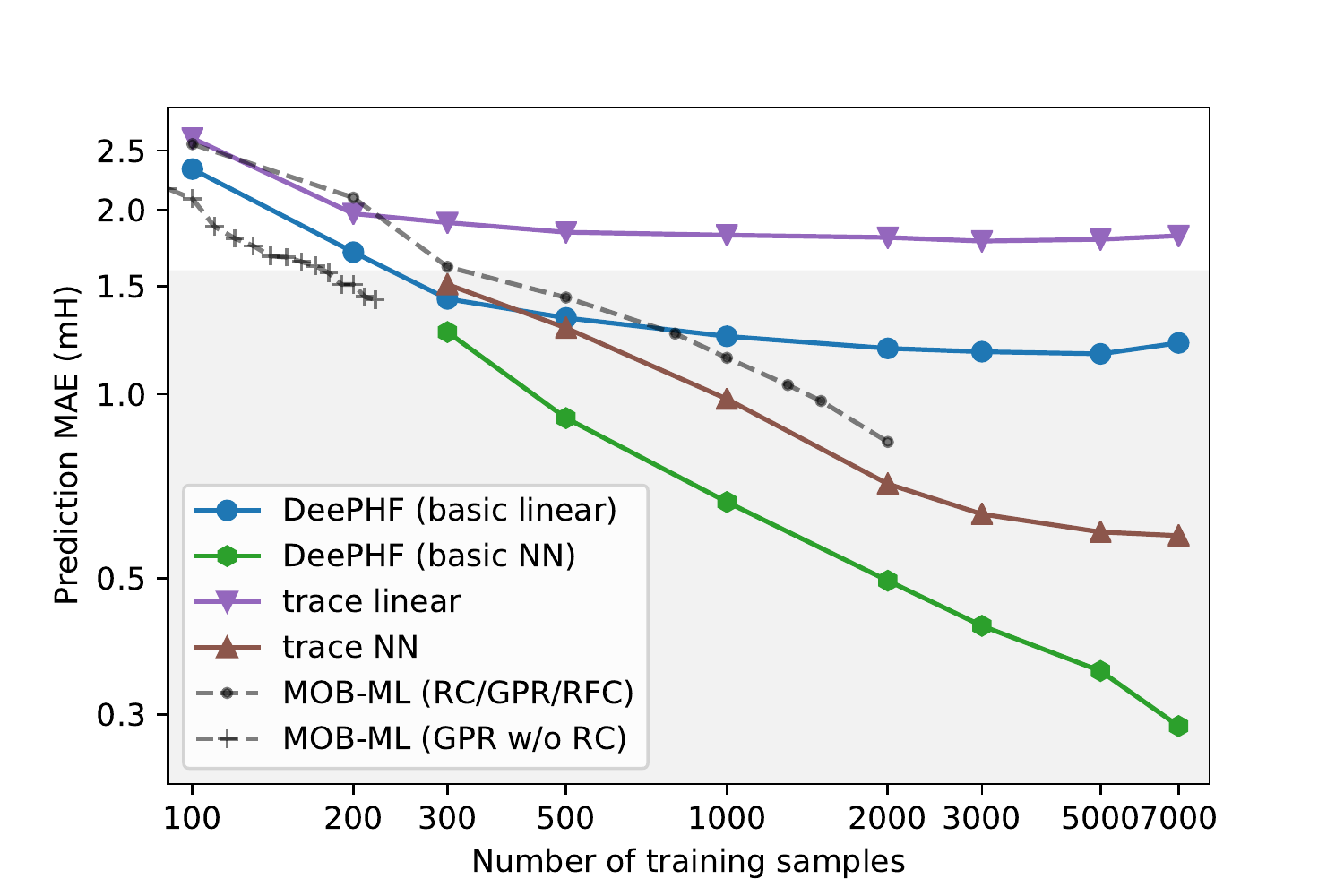}
    \caption{
    Learning curve of the proposed method on the QM7b-T dataset, using trace as descriptor. Results of using basic eigenvalue descriptor as well as results from Ref.~\citenum{cheng2019regression} are also included for comparison. Grey shaded area indicates the region where the error is smaller than chemical accuracy (1 kcal/mol).  
    }
    \label{fig:trqm7lc}
\end{figure}

Finally, we would  like to point out the eigenvalue construction in DeePHF  is rather important especially when we 
deal  with large datasets. 
As an example,  let us examine another commonly used way of imposing rotational symmetry,  summing over the the angular indices -- the trace construction, using $\mr{Tr} \left[\left(\cD^I_{nl}\right)_{mm'}\right] = \sum_m \left(\cD^I_{nl}\right)_{mm}$ as descriptors.  As shown in Fig~\ref{fig:trqm7lc}, such a construction 
performs much  worse than the eigenvalue construction using the same density matrix. Moreover, the testing error of the trace construction  saturates  when the number of training samples increases over 3000, even if we  use neural network fitting functions. Such a behavior suggests that the trace construction is reaching its limit of expressive power. 
By using eigenvalues as descriptors, DeePHF  has a more faithful representation of the ground state density matrix, and is able to capture the functional dependence of the correlation energies even for large datasets.

\subsection{The ANI-1ccx Dataset}
As mentioned previously, we find training on QM7b-T may be inadequate for us to acquire a model that can transfer well to larger molecules like those in GDB-13-T. Therefore, we take a step further to include the newly published ANI-1ccx dataset\cite{smith2020ani} as our training set. The ANI-1ccx dataset includes about 500k configurations of diverse molecule that are consisted by C, H, N and O. These configurations are intelligently selected from a larger set of data, ANI-1x\cite{smith2018less}, with around 5M configurations. The energies of these molecule configurations are calculated by the CCSD(T)*/CBS method\cite{smith2019approaching}, an accurate approximation of CCSD(T) extrapolating to the complete basis limit. This dataset is originally constructed and used to train a atom-based neural network potential model, also named ANI-1ccx\cite{smith2019approaching}, hence is designed to be diverse and unduplicated. 

Given this diverse nature of the dataset, we do not test the transferability within the dataset. Instead, we perform the same test used to examine the ANI-1ccx model\cite{smith2019approaching}. Due to the lack of testing data, we only measure the accuracy of hydrocarbon reaction and isomerization energies using HC7/11\cite{peverati2011generalized} and ISOL6\cite{luo2011validation} benchmarks. We examine the neural network model with basic descriptors. Detailed results can be find in Table~\ref{tab:ani}.

On both benchmarks, DeePHF performs pretty good and the predictions are within chemical accuracy. We note here the ANI-1cxx model is obtained by first training one a even larger dataset ANI-1x and adapting to ANI-1ccx data through transfer learning. Hence it in fact utilizes more information than DeePHF do. Also note that the CCSD(T)*/CBS row corresponds to the method that provides labels of DeePHF training. It might looks mysterious in the HC7/11 case that the DeePHF model even outperforms its "ground truth". This is may be understood that the DeePHF prediction happens to lies in the middle between CCSD(T)*/CBS method and the best estimation reference. Since the benchmark only contains six reaction, such behavior can happen by chance.

\begin{table}[htbp]
\centering
\caption{Mean absolute error of reaction and isomerization energies comparing to the best estimation reference. Results of methods other than DeePHF come from Ref.~\citenum{smith2019approaching}. Errors are given in mH.}
\begin{tabular}{l||l|l}
Methods     & HC7/11 & ISOL6 \\
\hline
$\omega$B97X/6-31g* & 26.1  & 6.1  \\
CCSD(T)*/CBS & 2.5   & 0.8  \\
ANI-1ccx     & 4.0   & 2.4  \\
\textbf{DeePHF} & 1.1   & 1.4  \\
\end{tabular}
\label{tab:ani}
\end{table}

\section{Conclusions}

We have developed the DeePHF scheme, a machine learning-based method for predicting
the  ground-state energy of electronic structures  with an accuracy comparable to post Hartree-Fock
methods like CCSD(T) and a computational cost comparable to that of HF and DFT. In this paper, we focused
on predicting the correlation energies using the ground state density matrix calculated by HF or DFT. 
The models are designed to satisfy several requirements, namely  universality, locality, symmetry, accuracy, and efficiency.

The procedure of the DeePHF scheme consists of two parts:  The selection of the descriptors and the construction
of the fitting function using the descriptors as the input. The descriptors are constructed by projecting the
density matrix onto some judiciously chosen local basis.
The fitting function is constructed with either a linear model or a neural network model, depending on the required accuracy.
An active learning procedure is introduced to minimize the number of data for labeling from a given unlabeled dataset.

We examined the performance of DeePHF on organic molecular systems using datasets provided in Ref.~\citenum{cheng2019data}. 
In all the tests, the accuracy of our model is either comparable to or better than previously reported results\cite{cheng2019regression,dick2020machine}. 
For simple datasets, we demonstrated that simple linear regression gives satisfactory results already. 
For the larger dataset QM7b-T, DeePHF performs better than Ref.~\citenum{cheng2019regression} in terms of accuracy and sample efficiency.

% \begin{acknowledgement}
\section{Acknowledgement}
The work of Y. C., L. Z. and W. E was supported in part by a gift from iFlytek to Princeton University, the ONR grant N00014-13-1-0338, and the Center Chemistry in Solution and at Interfaces (CSI) funded by the DOE Award DE-SC0019394.
The work of H. W. is supported by the National Science Foundation of China under Grant No. 11871110, the National Key Research and Development Program of China under Grants No. 2016YFB0201200 and No. 2016YFB0201203, and Beijing Academy of Artificial Intelligence (BAAI).
% \end{acknowledgement}

\appendix

\section{Projection Basis Coefficients}
\label{appendix:basis}
The basis coefficients are given in NWChem format that could be found in Basis Set Exchange website\cite{pritchard2019new}. Here we do not include the element name since it does not depend on the element type. \\
For the 108 basis functions the coefficients are as follows.
\begin{verbatim}
SPD
9.8526125336e+02  1.0 -1.0  0.0  0.0  0.0  0.0  0.0  0.0  0.0  0.0  0.0  0.0
1.9461950684e+02  0.0  1.0 -1.0  0.0  0.0  0.0  0.0  0.0  0.0  0.0  0.0  0.0
5.7665039062e+01  0.0  0.0  1.0 -1.0  0.0  0.0  0.0  0.0  0.0  0.0  0.0  0.0
1.7085937500e+01  0.0  0.0  0.0  1.0 -1.0  0.0  0.0  0.0  0.0  0.0  0.0  0.0
7.5937500000e+00  0.0  0.0  0.0  0.0  1.0 -1.0  0.0  0.0  0.0  0.0  0.0  0.0
3.3750000000e+00  0.0  0.0  0.0  0.0  0.0  1.0 -1.0  0.0  0.0  0.0  0.0  0.0
2.2500000000e+00  0.0  0.0  0.0  0.0  0.0  0.0  1.0 -1.0  0.0  0.0  0.0  0.0
1.5000000000e+00  0.0  0.0  0.0  0.0  0.0  0.0  0.0  1.0 -1.0  0.0  0.0  0.0
1.0000000000e+00  0.0  0.0  0.0  0.0  0.0  0.0  0.0  0.0  1.0 -1.0  0.0  0.0
6.6666666667e-01  0.0  0.0  0.0  0.0  0.0  0.0  0.0  0.0  0.0  1.0 -1.0  0.0
4.4444444444e-01  0.0  0.0  0.0  0.0  0.0  0.0  0.0  0.0  0.0  0.0  1.0 -1.0
2.9629629630e-01  0.0  0.0  0.0  0.0  0.0  0.0  0.0  0.0  0.0  0.0  0.0  1.0
\end{verbatim}
For the 81 basis functions the coefficients are as follows.
\begin{verbatim}
SPD
6.400000e+01  1.0 -1.0  0.0  0.0  0.0  0.0  0.0  0.0  0.0
3.200000e+01  0.0  1.0 -1.0  0.0  0.0  0.0  0.0  0.0  0.0
1.600000e+01  0.0  0.0  1.0 -1.0  0.0  0.0  0.0  0.0  0.0
8.000000e+00  0.0  0.0  0.0  1.0 -1.0  0.0  0.0  0.0  0.0
4.000000e+00  0.0  0.0  0.0  0.0  1.0 -1.0  0.0  0.0  0.0
2.000000e+00  0.0  0.0  0.0  0.0  0.0  1.0 -1.0  0.0  0.0
1.000000e+00  0.0  0.0  0.0  0.0  0.0  0.0  1.0 -1.0  0.0
5.000000e-01  0.0  0.0  0.0  0.0  0.0  0.0  0.0  1.0 -1.0
2.500000e-01  0.0  0.0  0.0  0.0  0.0  0.0  0.0  0.0  1.0
\end{verbatim}

\bibliographystyle{plain} % comment out for jpc
\bibliography{ref} 

%%% for jpc toc image %%%
% \newpage
% \begin{figure}[h]
%     \centering
%     \includegraphics[width=8.25cm]{img/log_sample_curve.pdf}
%     \label{TOC Graphic}
% \end{figure}

\end{document}